\documentclass{article}

\usepackage{PRIMEarxiv}
\usepackage{multirow}
\usepackage[utf8]{inputenc} 
\usepackage[T1]{fontenc}    
\usepackage{hyperref}       
\usepackage{url}            
\usepackage{booktabs}       
\usepackage{amsfonts}       
\usepackage{nicefrac}       
\usepackage{microtype}      
\usepackage{lipsum}
\usepackage{fancyhdr}       
\usepackage{graphicx}       
\graphicspath{{media/}}
\usepackage{amsmath}
\usepackage[table,xcdraw]{xcolor}

\title{WaveMixSR-V2: Enhancing Super-resolution with Higher Efficiency
}

\author{
  Pranav Jeevan\thanks{Equal Contribution.} \hspace{10mm} Neeraj Nixon\footnotemark[1] \hspace{10mm} Amit Sethi \\
  Department of Electrical Engineering \\
  Indian Institute of Technology Bombay \\
  Mumbai, India  \\
  \texttt{\{pjeevan, 20d070056, asethi\}@iitb.ac.in} \\
}

\begin{document}
\maketitle

\begin{abstract}
Recent advancements in single image super-resolution have been predominantly driven by token mixers and transformer architectures. WaveMixSR utilized the WaveMix architecture, employing a two-dimensional discrete wavelet transform for spatial token mixing, achieving superior performance in super-resolution tasks with remarkable resource efficiency. In this work, we present an enhanced version of the WaveMixSR architecture by (1) replacing the traditional transpose convolution layer with a pixel shuffle operation and (2) implementing a multistage design for higher resolution tasks ($4\times$). Our experiments demonstrate that our enhanced model -- WaveMixSR-V2 -- outperforms other architectures in multiple super-resolution tasks, achieving state-of-the-art for the BSD100 dataset, while also consuming fewer resources, exhibits higher parameter efficiency, lower latency and higher throughput. Our code is available at \url{https://github.com/pranavphoenix/WaveMixSR}
\end{abstract}


\begin{figure}[h!]
\centering
\includegraphics[scale=0.52]{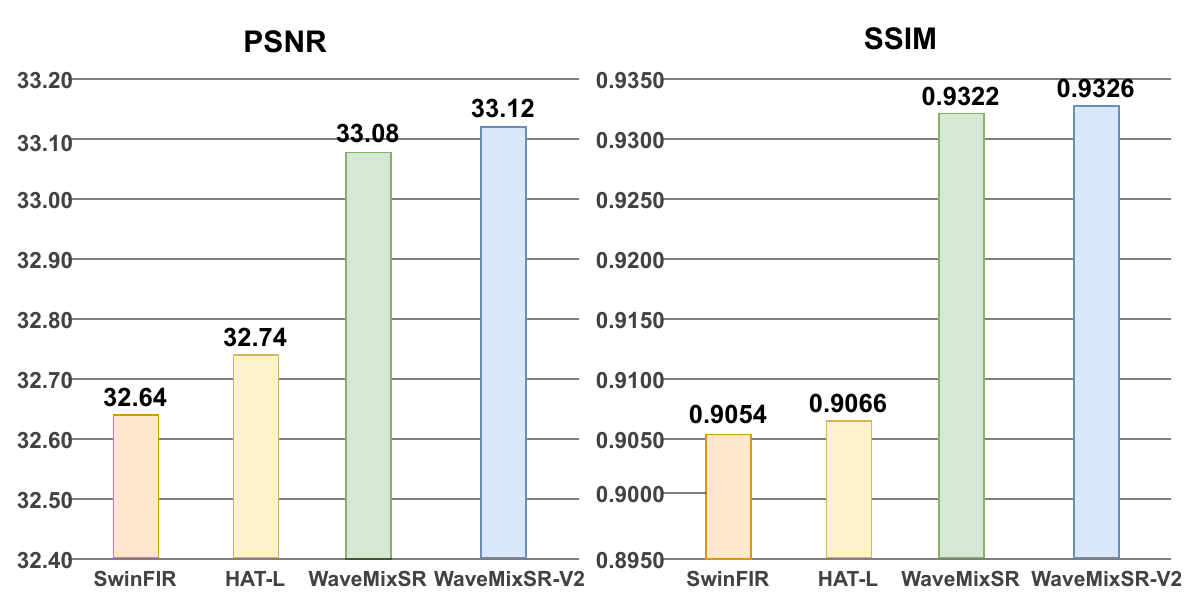} 
\caption{Comparison of PSNR and SSIM for $2\times$ SR on BSD100 dataset shows WaveMixSR-V2 surpasses the previous
state-of-the-art WaveMixSR and other methods such as HAT and SwinFIR. $4\times$ SR results in Appendix.}
\label{fig2}
\vspace{-2mm}
\end{figure}

\section{Introduction}

Single-image super-resolution (SISR) is a key task in image reconstruction, aiming to transform low-resolution (LR) images into high-resolution (HR) by predicting and restoring missing details. This process requires capturing both local information and global context. Recent advancements in super-resolution, particularly with attention-based transformers like SwinFIR~\cite{zhang2023swinfir} and hybrid attention transformer~\cite{chen2023activating}, have surpassed traditional CNN approaches due to their ability to capture long-range dependencies. However, transformers face challenges with quadratic complexity in self-attention, leading to high resource demands and requiring large datasets. To overcome this, token-mixer models such as WaveMixSR~\cite{Jeevan_2024_WACV}, which uses a two-dimensional discrete wavelet transform, have shown potential for improved efficiency and even superior performance. Building on the strengths of WaveMixSR, we propose enhancements to the model by rethinking its upsampling strategy inside the WaveMix blocks and changing the single stage design. Details of architecture, more results, including those on $4\times$ SR, and ablation studies are provided in the Appendix.

\section{Architectural Improvements}

\subsection{Multi-stage Design}

We have made significant improvements to the WaveMixSR model, focusing on two key aspects. First, we addressed how the model handles SR tasks higher than $2\times$. In the original WaveMixSR~\cite{Jeevan_2024_WACV}, all SR tasks were performed by directly resizing the LR image to HR using a single upsampling layer. This layer relied on non-parametric upsampling techniques, such as bilinear or bicubic interpolation, which limited the model's ability to fine-tune and optimize the SR process across different scales. Our approach involved transitioning from this single-stage design to a more robust multi-stage design. In our new architecture, we introduced a series of resolution-doubling $2\times$ SR blocks, which progressively doubles the resolution step by step. This multi-stage approach allows for better SR performance at higher scales while reducing resource consumption.

\begin{figure}[h!]
\centering
\includegraphics[scale=0.9]{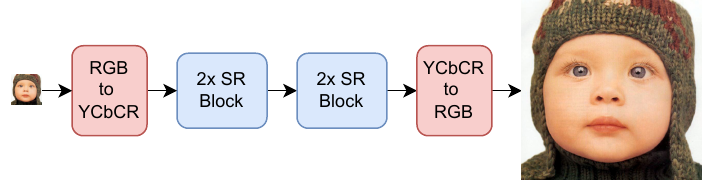} 
\caption{Architecture of WaveMixSR-V2 showing $4\times$ SR with two $2\times$ SR blocks in series. Details in Appendix.}
\label{fig3}

\end{figure}

For instance, in an $4\times$ super-resolution task, instead of directly upsampling the LR image to HR using a single interpolation layer, the model now proceeds through a series of two $2\times$ SR blocks as shown in Fig.~\ref{fig3} By incrementally increasing the resolution (doubling in each stage), the model is better able to refine the details at each step, leading to superior super-resolution performance compared to the single upsampling operation used in the original WaveMixSR.

\subsection{PixelShuffle}

We introduce a key modification to the WaveMixSR model by replacing the transposed convolution operation in the WaveMix blocks with a PixelShuffle~\cite{shi2016realtimesingleimagevideo} operation followed by a convolution layer (WaveMixSR-V2 block) as shown in Fig.~\ref{fig2}. While the original WaveMixSR used transposed convolutions, which involved numerous parameters and high computational cost, PixelShuffle upsamples the image more efficiently by rearranging pixels from feature maps. This significantly reduces the number of parameters, enhancing the model's efficiency. The subsequent convolution layer after PixelShuffle allows the model to continue learning and refining features effectively. Moreover, PixelShuffle avoids the checkerboard artifacts commonly introduced by transposed convolutions, producing smoother and more natural-looking images while maintaining high-quality super-resolution outputs.

Incorporating these improvements in the architecture has enabled WaveMixSR-V2 to achieve new state-of-the-art (SOTA) performance on the BSD100 dataset~\cite{martin2001database}. Notably, it accomplishes this with less than half the number of parameters, lesser computations and lower latency compared to WaveMixSR (previous SOTA).

\begin{figure}[h!]
\centering
\includegraphics[scale=0.75]{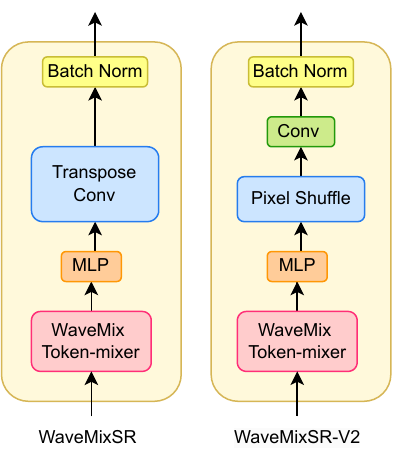} 
\caption{Simplified block diagram of WaveMix block in WaveMixSR (on the left) and WaveMixSR-V2 block (on the right). Details in Appendix.}
\label{fig2}

\end{figure}



\begin{table}[h!]
\centering

\begin{tabular}{l r r }
\toprule
Model   & \#Params. & \#Multi-Adds. \\ \midrule
SwinIR~\cite{liang2021swinir}      & 11.8 M             & 49.6 G                 \\ 
HAT~\cite{chen2023activating}        & 20.8 M             & 103.7 G                \\ 
WaveMixSR~\cite{Jeevan_2024_WACV} & 1.7 M    & 25.8 G        \\ 
\textbf{WaveMixSR-V2} & \textbf{0.7 M}     & \textbf{25.6 G}        \\ \bottomrule
\end{tabular}
\vspace{2mm}
\caption{Model complexity comparison of WaveMixSR-V2 with other state-of-the-art methods such as WaveMixSR, SwinIR and HAT on 4$\times$ SR of 64$\times$64 input patch.}
\label{table:model_comparison}

\end{table}

\bibliographystyle{unsrt}  
\bibliography{references}  

\appendix

\begin{figure*}[t]
\centering
\includegraphics[scale=0.65]{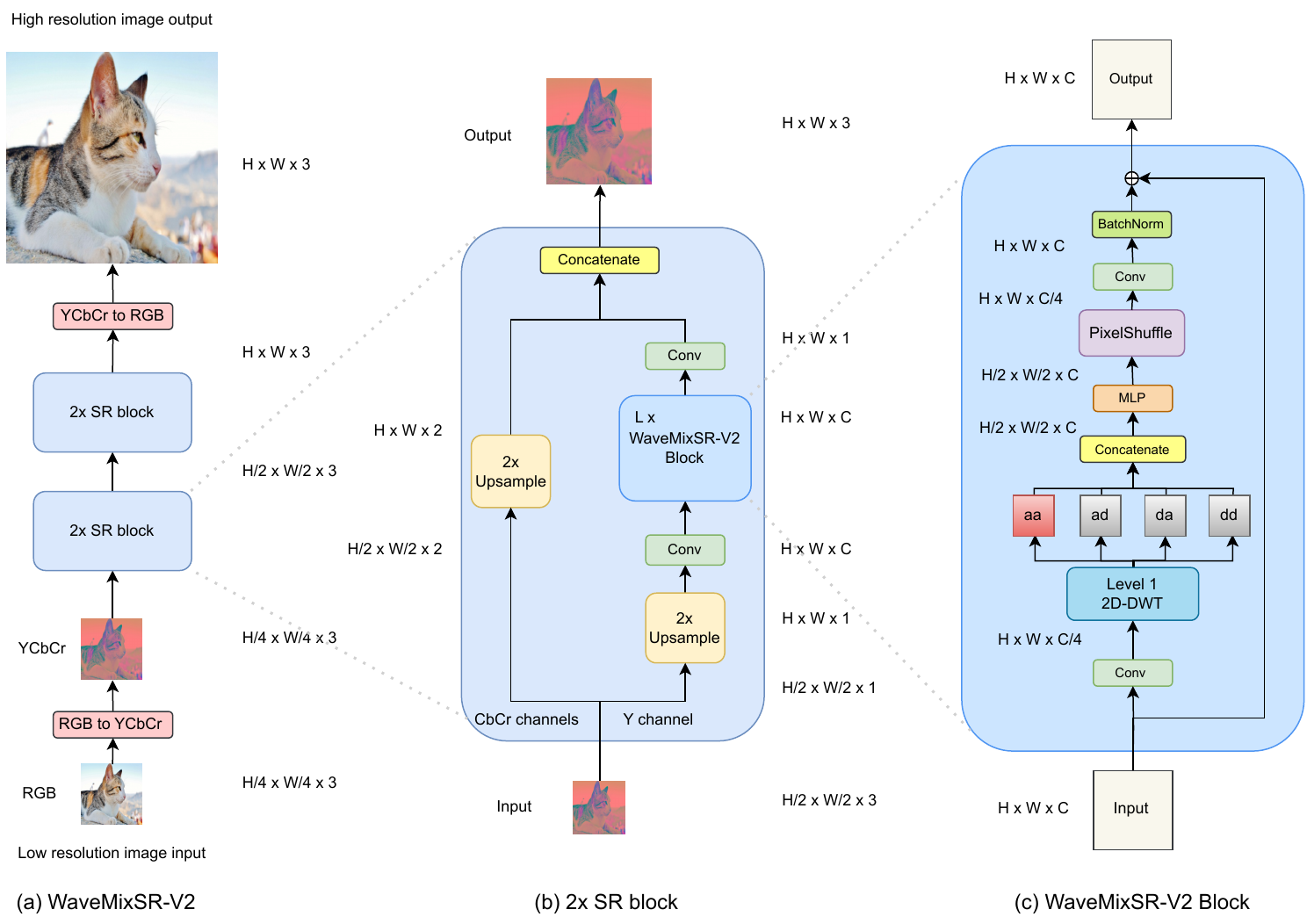} 
\caption{Architecture of WaveMixSR-V2. (a) The application of WaveMixSR-V2 for $4\times$ SR is shown featuring two $2\times$ SR blocks stacked in series. For higher SR tasks, more $2\times$ SR blocks will be added.  (b) The details of the $2\times$ SR block and (c) shows the WaveMixSR-V2 block that replaces the transposed convolution with a PixelShuffle operation followed by a convolution.} 

\label{fig1} 
\end{figure*}

\section{Architecture}

\subsection{WaveMixSR-V2 Block}

The core part of our super-resolution (SR) architecture are the WaveMixSR-V2 blocks shown in Fig.~\ref{fig1}(c). We created these by modifying the WaveMix~\cite{jeevan2023wavemix} blocks used in WaveMixSR model~\cite{Jeevan_2024_WACV}. We replace the transposed convolution used in WaveMix block with a PixelShuffle~\cite{shi2016realtimesingleimagevideo} operation followed by a convolutional layer.

Denoting input and output tensors of the WaveMixSR-V2 block by $\textbf{x}_{in}$ and $\textbf{x}_{out}$, respectively; the four wavelet filters along with their downsampling operations at each level by $w_{aa},w_{ad},w_{da},w_{dd}$ ($a$ for approximation, $d$ for detail); convolution, multi-layer perceptron (MLP), PixelShuffle, and batch normalization operations by $c$, $m$, $p$, and $b$, respectively; and their respective trainable parameter sets by $\xi$, $\theta$, $\phi$, and $\gamma$, respectively; concatenation along the channel dimension by $\oplus$, and point-wise addition by $+$, the operations inside a WaveMixSR-V2 block can be expressed using the following equations:

\begin{equation} \label{eq:1}
    \textbf{x}_0 = c(\textbf{x}_{in},\xi);  \hspace{5mm}    
    \textbf{x}_{in}\in\mathbb{R}^{H\times W \times C}, \textbf{x}_0\in\mathbb{R}^{H\times W \times C/4}    
\end{equation}

\begin{equation}\label{eq:2}
    \textbf{x} = [w_{aa}(\textbf{x}_0) \oplus w_{ad}(\textbf{x}_0) \oplus w_{da}(\textbf{x}_0) \oplus w_{dd}(\textbf{x}_0)]; \hspace{5mm}
    \textbf{x}\in\mathbb{R}^{H/2\times W/2 \times 4C/4}
\end{equation}

\begin{equation} \label{eq:3}
    \tilde{\textbf{x}} = b(c(p(m(\textbf{x},\theta),\phi),\xi),\gamma); \hspace{5mm}\tilde{\textbf{x}}\in\mathbb{R}^{H\times W \times C}      
\end{equation}

\begin{equation}\label{eq:4}
    \textbf{x}_{out} = \tilde{\textbf{x}}_1 + \textbf{x}_{in}; \hspace{5mm}\hspace{2cm}\textbf{x}_{out} \in\mathbb{R}^{H\times W \times C} 
\end{equation}

\begin{table*}[]
\centering
\begin{tabular}{@{}llrrrrrr@{}}
\toprule
\multirow{3}{*}{Model} & \multirow{3}{*}{Training dataset} & \multicolumn{4}{c}{Testing metrics on BSD100} \\ 
\cmidrule{3-6}  
 &  & \multicolumn{2}{c}{$2\times$ SR} & \multicolumn{2}{c}{$4\times$ SR} \\ \cmidrule{3-6} 
 &  & PSNR & SSIM & PSNR & SSIM \\ \midrule
EDSR~\cite{lim2017enhanced} & DIV2K & 32.32 & 0.9013  & 27.71 & 0.7420 \\
RCAN~\cite{zhang2018image} & DIV2K & 32.41 & 0.9027  & 27.77 & 0.7436 \\
SAN~\cite{Chen_2023_CVPR} & DIV2K & 32.42 & 0.9028 &  27.78 & 0.7436 \\
IGNN~\cite{Chen_2023_CVPR} & DIV2K & 32.41 & 0.9025 &  27.77 & 0.7434 \\
HAN~\cite{niu2020single} & DIV2K & 32.41 & 0.9027 &  27.80 & 0.7442 \\
NLSN~\cite{Chen_2023_CVPR} & DIV2K & 32.43 & 0.9027 &  27.78 & 0.7444 \\
RCN-it~\cite{Chen_2023_CVPR} & DF2K & 32.48 & 0.9034 &  27.87 & 0.7459 \\
SwinIR~\cite{liang2021swinir} & DF2K & 32.53 & 0.9041 &  27.92 & 0.7489 \\
EDT~\cite{li2022efficient} & DF2K & 32.52 & 0.9041 &  27.91 & 0.7483 \\
HAT~\cite{Chen_2023_CVPR} & DF2K & 32.62 & 0.9053 &  28.00 & 0.7517 \\
SwinFIR*~\cite{zhang2023swinfir} & DF2K & 32.64 & 0.9054 &  28.03& 0.7520 \\
HAT-L*~\cite{Chen_2023_CVPR} & DF2K & 32.74 & 0.9066  & \textbf{28.09} & 0.7551 \\
WaveMixSR~\cite{Jeevan_2024_WACV} & DIV2K & 33.08 & 0.9322 &  27.65 & 0.7605 \\
\textbf{WaveMixSR-V2} & DIV2K & \textbf{33.12} & \textbf{0.9326} &  27.87 & \textbf{0.7640} \\
\bottomrule
\end{tabular}
\caption{Quantitative comparison with previous state-of-the-art methods on the BSD100 dataset shows that WaveMixSR-V2 performs better using less training data (* indicates models that were pre-trained on ImageNet).}
\label{tab:bsd100}
\end{table*}

\begin{figure}[]
\centering
\includegraphics[width=0.5\textwidth]{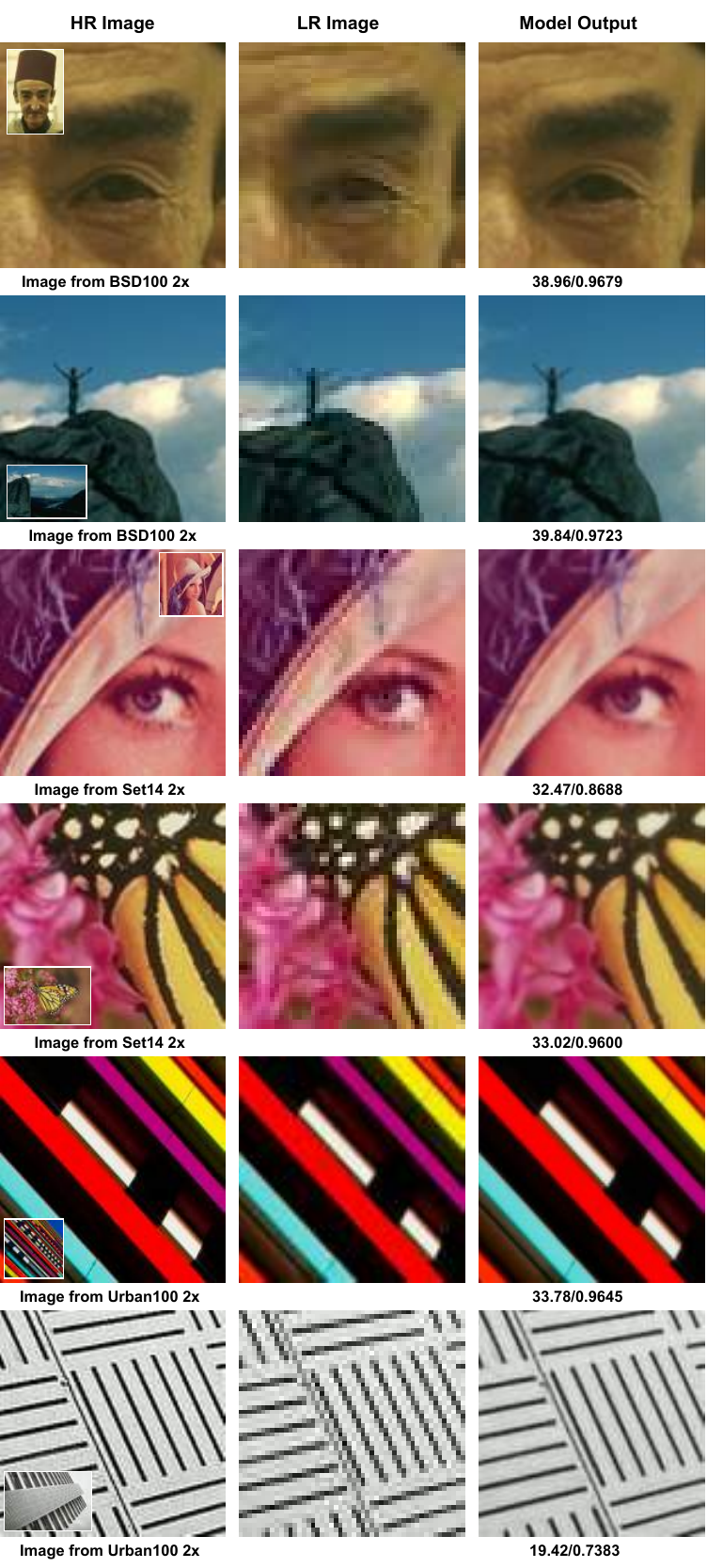} 
\caption{Visual results of $2\times$ SR on BSD100 dataset. Each column from the left shows a patch from the HR image (shown as a small image near the corner), the same patch extracted from the LR image, and a patch taken from the model output respectively. The filename of the image is given below the HR image and the PSNR/SSIM of the model output is reported at below the model output. The values displayed are computed for the whole image and not just the patch.}
\label{fig2}
\end{figure}

\begin{table}[]
\centering

\begin{tabular}{@{}lrrrr@{}}
\toprule
Model        & \begin{tabular}[c]{@{}r@{}}Training\\ Latency \textdownarrow\\ (ms)\end{tabular} & \begin{tabular}[c]{@{}r@{}}Training\\ Throughput\textuparrow\\ (fps)\end{tabular} & \begin{tabular}[c]{@{}r@{}}Inference\\ Latency \textdownarrow\\ (ms)\end{tabular} & \begin{tabular}[c]{@{}r@{}}Inference\\ Throughput \textuparrow\\ (fps)\end{tabular} \\ \midrule
WaveMixSR    & 22.8                                                              & 43.8                                                                  & 18.6                                                                     & 53.7                                                                   \\
WaveMixSR-V2 & 19.6                                                              & 50.8                                                                  & 12.1                                                                     & 82.6                                                                   \\ \bottomrule
\end{tabular}%
\vspace{2mm}
\caption{Comparison of latency and throughput of WaveMixSR-V2 and WaveMixSR shows that WaveMixSR-V2 is significantly faster than WaveMixSR}
\label{tab:my-table1}
\end{table}

\begin{table}[]
\centering
\begin{tabular}{@{}lrrllrr@{}}
\toprule
\multirow{2}{*}{Scale}  & \multicolumn{2}{c}{$2\times$} & \multicolumn{2}{c}{$4\times$}  \\ \cmidrule{2-5} 
  & PSNR & SSIM & PSNR & SSIM \\ \midrule
Set5  & 35.85 & 0.9522 & 29.49 & 0.8633 \\

Set14  & 31.20 & 0.9000 & 26.39 & 0.7521 \\

Urban100  & 29.20 & 0.9076 & 23.93 & 0.7384
 \\ \bottomrule
\end{tabular}
\vspace{2mm}
\caption{Quantitative results of WaveMixSR-V2 on other benchmark SR datasets}
\label{tab:data}
\end{table}

The WaveMixSR-V2 block extracts learnable and space-invariant features using a convolutional layer, followed by spatial token-mixing and downsampling for scale-invariant feature extraction using 2 dimenional-discrete wavelet transform (2D-DWT)~\cite{pyTorchWavelets}, followed by channel-mixing using a learnable MLP (1$\times$1 conv) layer, followed by restoring spatial resolution of the feature map using PixelShuffle operation. The use of trainable convolutions \emph{before} the wavelet transform allows the extraction of only those feature maps that are suitable for the chosen wavelet basis functions. The convolutional layer $c$ decreases the embedding dimension  $C$ by a factor of four so that the concatenated output $\textbf{x}$ after 2D-DWT has the same number of channels as the input $\textbf{x}_{in}$ (Eq.\ref{eq:1} and Eq.\ref{eq:2}). That is since 2D-DWT is a lossless transform, it expands the number of channels by the same factor (using concatenation) by which it reduces the spatial resolution by computing an approximation sub-band (low-resolution approximation) and three detail sub-bands (spatial derivatives)~\cite{57199} for each input channel (Eq.\ref{eq:2}). The use of this image-appropriate and lossless downsampling using 2D-DWT allows WaveMixSR-V2 to use fewer layers and parameters.

The output $\hat{\textbf{x}}$ is then passed to an MLP layer $m$, which has two $1\times1$ convolutional layers with an inverse bottleneck design (multiplication factor $> 1$) separated by a GELU non-linearity. After this, the feature map resolution is doubled using PixelShuffle operation $p$. Since PixelShuffle reduces the channel dimension by 4, we use another convolution layer $c$ to increase the channel dimension back to $C$. This is followed by 
batch normalization $b$ (Eq.\ref{eq:3}). A residual connection is used to ease the flow of the gradient~\cite{he2015deep} (Eq.\ref{eq:4}). The WaveMixSR-V2 block ensures that input and output resolution are the same.

Among the different types of mother wavelets available, we used the Haar wavelet (a special case of the Daubechies wavelet~\cite{57199}, also known as Db1), which is frequently used due to its simplicity and faster computation. Haar wavelet is both orthogonal and symmetric in nature and has been extensively used to extract basic structural information from images~\cite{Porwik2004TheHT}. For even-sized images, it reduces the dimensions exactly by a factor of $2$, which simplifies the designing of the subsequent layers.

\subsection{$2\times$ SR Block}

As shown in Fig.~\ref{fig1}(b), the $2\times$ SR block has two paths -- one for handling the Y channel and another for the CbCr channels of the input image. The Y channel is used for the path with learning using WaveMixSR-V2 blocks because the Y channel contains most of the image details and is less affected by color changes. It first upsamples the image to HR size using a parameter-free upsampling block using bilinear or bicubic interpolation. The output of upsampling block, is sent to a convolutional layer to increase the number of feature maps before sending it to the WaveMixSR-V2 blocks. We connected $L$ WaveMixSR-V2 blocks in series to create high-resolution feature maps. The output from the final WaveMixSR-V2 blocks is then passed through a convolutional layer which reduces the channel dimension and returns a single channel output. 

The second parallel path takes the two CbCr channels and  passes it through an upsampling layer where the resolution is doubled. This HR CbCr channel is concatenated with the Y-channel output from the first path, thereby creating the 3-channel YCbCr HR output, which is converted to RGB to obtain the doubled resolution output image.

\subsection{WaveMixSR-V2}

The LR input image in RGB space is first converted to YCbCr space before sending to the model as shown Fig.~\ref{fig1}(a). It is then passed through a series of $2\times$ SR blocks. For $2\times$ SR, we use just one $2\times$ SR block. For higher SR tasks, we can modify the network by adding as many  $2\times$ SR blocks as required to achieve as much SR as needed, as adding one $2\times$ SR block doubles the resolution. Finally, the output from $2\times$ SR blocks are converted back to RGB space to get final output.

\section{Implementation Details}

We used DIV2K dataset~\cite{8014884} for training WaveMixSR-V2. We did not employ any pre-training on larger datasets such as DF2K~\cite{8014884} or ImageNet~\cite{5206848} to compare the performance in training data-constrained settings. The performance of WaveMixSR-V2 was tested on four benchmark datasets -- BSD100~\cite{937655}, Urban100~\cite{7299156}, Set5~\cite{BMVC.26.135}, and Set14~\cite{10.1007/978-3-642-27413-8_47}, .

All experiments were done with a single 48 GB Nvidia A6000 GPU. We used AdamW optimizer ($\alpha = 0.001, \beta_{1} = 0.9, \beta_{2}=0.999, \epsilon = 10^{-8}$) with a weight decay of 0.01 during initial epochs and then used SGD with a learning rate of $0.001$ and momentum $= 0.9$ during the final 50 epochs~\cite{keskar2017improving, https://doi.org/10.48550/arxiv.2201.10271}. A dropout of 0.3 is used in our experiments. A batch size of 1 was used when the full-resolution images were passed to the model and a batch size of 432 was used when images were passed as $64\times64$ resolution patches.

The LR images were generated from the HR images by using bicubic down-sampling in Pytorch. We used the full-resolution HR image as the target and generated the input LR image using down-sampling for each of the SR tasks. No data augmentations were used while training the WaveMixSR-V2 models. Huber loss was used to optimize the parameters. We used automatic mixed precision in PyTorch during training. For the quantitative results, PSNR and SSIM (calculated on the Y channel) are reported. 

 The embedding dimension of 144 was used in WaveMixSR-V2 blocks. The convolutions layers before and after the WaveMixSR-V2 blocks which were used to vary channel dimensions employed $3\times3$ kernels with stride and padding set to 1 to maintain the feature resolution.

\begin{table*}[t]
\centering
\resizebox{\textwidth}{!}{%
\begin{tabular}{@{}rrrrrcccccccc@{}}
\toprule
\multirow{2}{*}{Input} & \multirow{2}{*}{ Pixel}  & \multirow{2}{*}{Content } & \multirow{2}{*}{Adversarial}  & \multirow{2}{*}{Noise} & \multicolumn{2}{c}{BSD100}         & \multicolumn{2}{c}{Set5}           & \multicolumn{2}{c}{Set14}          & \multicolumn{2}{c}{Urban100}       \\ \cmidrule{6-13} 
          Resolution                  &     Loss $\lambda_0$                    &    Loss $\lambda_1$                     &      Loss $\lambda_2$                 & Added& \multicolumn{1}{c}{PSNR}  & SSIM   & \multicolumn{1}{c}{PSNR}  & SSIM   & \multicolumn{1}{c}{PSNR}  & SSIM   & \multicolumn{1}{c}{PSNR}  & SSIM   \\ \midrule
\multicolumn{1}{r}{$128\times128$}     & \multicolumn{1}{r}{1}     & \multicolumn{1}{r}{0.01}     & \multicolumn{1}{r}{0.001} & yes                    & \multicolumn{1}{c}{28.39} & 0.8816 & \multicolumn{1}{c}{30.16} & 0.8996 & \multicolumn{1}{c}{27.76} & 0.8667 & \multicolumn{1}{c}{25.14} & 0.8520 \\
\multicolumn{1}{r}{$64\times64$}     &\multicolumn{1}{r}{1}     & \multicolumn{1}{r}{0.01}     & \multicolumn{1}{r}{0.001} & yes                    & \multicolumn{1}{c}{28.62} & 0.8661 & \multicolumn{1}{c}{29.96} & 0.8728 & \multicolumn{1}{c}{27.67} & 0.8421 & \multicolumn{1}{c}{25.24} & 0.8368 \\
\multicolumn{1}{r}{$128\times128$}     &\multicolumn{1}{r}{1}     & \multicolumn{1}{r}{0}        & \multicolumn{1}{r}{0.01}  & yes                    & \multicolumn{1}{c}{30.02} & 0.9065 & \multicolumn{1}{c}{31.01} & 0.9146 & \multicolumn{1}{c}{28.46} & 0.8756 & \multicolumn{1}{c}{25.78} & 0.8584 \\ 
\multicolumn{1}{r}{$64\times64$}     &\multicolumn{1}{r}{1}     & \multicolumn{1}{r}{0}        & \multicolumn{1}{r}{0.01}  & yes                    & \multicolumn{1}{c}{30.23} & 0.9023 & \multicolumn{1}{c}{30.84} & 0.9121 & \multicolumn{1}{c}{28.41} & 0.8154 & \multicolumn{1}{c}{25.93} & 0.8434 \\ 
\multicolumn{1}{r}{$128\times128$}     &\multicolumn{1}{r}{1}     & \multicolumn{1}{r}{0.01}     & \multicolumn{1}{r}{0.001} & no                     & \multicolumn{1}{c}{29.36} & 0.9008 & \multicolumn{1}{c}{30.46} & 0.9064 & \multicolumn{1}{c}{27.58} & 0.8660 & \multicolumn{1}{c}{25.02} & 0.8476 \\ 
\multicolumn{1}{r}{$64\times64$}     &\multicolumn{1}{r}{1}     & \multicolumn{1}{r}{0.01}     & \multicolumn{1}{r}{0.001} & no                     & \multicolumn{1}{c}{28.09} & 0.8815 & \multicolumn{1}{c}{29.48} & 0.8918 & \multicolumn{1}{c}{27.34} & 0.8664 & \multicolumn{1}{c}{24.83} & 0.8474 \\ 
\multicolumn{1}{r}{$128\times128$}     &\multicolumn{1}{r}{1}     & \multicolumn{1}{r}{0}     & \multicolumn{1}{r}{0} & no                     & \multicolumn{1}{c}{31.24} & 0.9267 & \multicolumn{1}{c}{33.12} & 0.9398 & \multicolumn{1}{c}{29.44} & 0.8957 & \multicolumn{1}{c}{26.49} & 0.8780 \\ 
\multicolumn{1}{r}{$64\times64$}     &\multicolumn{1}{r}{1}     & \multicolumn{1}{r}{0}     & \multicolumn{1}{r}{0} & no                     & \multicolumn{1}{c}{31.19} & 0.9263 & \multicolumn{1}{c}{32.91} & 0.9378 & \multicolumn{1}{c}{29.38} & 0.8948 & \multicolumn{1}{c}{26.46} & 0.8772 \\ \bottomrule
\end{tabular}
}
\caption{Quantitative comparison of WaveMixSR-V2 on benchmark datasets when using different combination of losses for various input resolutions.}
\label{tab:my-table}
\end{table*}

\begin{table*}[]
\centering
\begin{tabular}{@{}rrrrllrrllrr@{}}
\toprule
\multirow{2}{*}{Channel} & \multirow{2}{*}{Layer}  & \multicolumn{2}{c}{BSD100} & \multicolumn{2}{c}{Set5} & \multicolumn{2}{c}{Set14} & \multicolumn{2}{c}{Urban100} \\ \cmidrule{3-10}
 Dimension & Depth & PSNR & SSIM & PSNR & SSIM & PSNR & SSIM & PSNR & SSIM \\ \midrule
144 & 4 & 31.16 &	0.9267 &	33.00 &	0.9401 &	29.38 &	0.8964 &	26.45 &	0.8785  \\
160 & 4 & 31.16 &	0.9267 &	32.97 &	0.9400 &	29.38 &	0.8960 &	26.46 &	0.8786  \\
144 & 6 & 31.19 &	0.9263 &	32.91 &	0.9378 &	29.38 &	0.8948 &	26.46 &	0.8772 \\

\bottomrule
\end{tabular}
\caption{Results of ablation studies showing performance when we vary the channel dimension and layers of WaveMixSR-V2 blocks.}
\label{tab:ablation}
\end{table*}

\section{Results}

From Table~\ref{tab:bsd100}, it is evident that WaveMixSR-V2 has achieved state-of-the-art (SOTA) performance in $2\times$ and $4\times$ SR tasks on the BSD100 dataset. WaveMixSR-V2 attains this superior performance while utilizing significantly smaller DIV2K training data compared to other models, which typically rely on the much larger DF2K and ImageNet datasets. Even in the PSNR metric for $4\times$ SR, WaveMixSR-V2 is SOTA among all models trained solely on the smaller DIV2K dataset. In contrast, all models that outperform WaveMixSR-V2 in $4\times$ SR PSNR have been trained on the much larger DF2K data and even have leveraged ImageNet pre-training, further underscoring WaveMixSR-V2's efficiency in achieving SOTA results with fewer training data.

Table~\ref{tab:my-table1} provides a comparison of latency and throughput for WaveMixSR-V2 compared to its predecessor, WaveMixSR. WaveMixSR had previously set the benchmark for efficiency, known for its fast output, low GPU consumption, and parametric efficiency in SR tasks. However, WaveMixSR-V2 surpasses its predecessor with faster training and inference speeds. As shown in the Table ~\ref{tab:my-table1}, WaveMixSR-V2 exhibits a much lower latency and significantly higher throughput, both in training ($\sim15\%$ improvement) and inference ($~54\%$ improvement), cementing its status as one of the most efficient models for super-resolution.

In Table 3, we present the quantitative metrics for the results of $2\times$ and $4\times$ SR across various benchmark datasets, such as Set5, Set14, and Urban100. These datasets are widely used for evaluating super-resolution models. The results demonstrate that WaveMixSR-V2 performs competitively, delivering excellent outcomes in both $2\times$ and $4\times$ SR tasks across all datasets.

\section{Ablation Studies}
\subsection{WaveMixSR-V2 GAN}

We conducted experiments to check the performance of the WaveMixSR-V2 architecture when trained using a Generative Adversarial Network (GAN) framework, incorporating a relativistic discriminator~\cite{jolicoeurmartineau2018relativisticdiscriminatorkeyelement} with residual connections. We also experimented the impact of introducing Gaussian noise as an extra input channel alongside the RGB channels. The hypothesis was that this noise might enhance the HR quality by encouraging the model to incorporate higher frequency components. The experiments used images from the DIV2K dataset resized to different dimensions for training and were evaluated on benchmarks datasets.

\subsubsection{Loss}
In terms of loss functions, we used a combination of pixel loss ($L_{pixel}$), content loss ($L_{content}$)~\cite{ledig2017photorealistic}, and adversarial loss. The pixel loss was based on the Peak Signal-to-Noise Ratio (PSNR) to directly optimize the model for higher PSNR values. For content loss, a pre-trained VGG19 model was used as the feature extractor. Specifically, features were extracted from the 5th convolutional layer before the 4th max-pooling layer of the VGG-19 model. The adversarial loss was computed using a binary cross-entropy loss with logits on the output of the discriminator, comparing the predictions for real and generated images. The adversarial loss was to guide the generator to produce more realistic images, improving the fidelity of higher-frequency details.

The overall loss function combined pixel loss, content loss, and adversarial loss as follows:

\begin{center}
    \text{$Loss$} = $\lambda_0$\text{$L_{pixel}$} + $\lambda_1$ \text{$L_{content}$} + $\lambda_2$ \text{$L_{adversarial}$} 
\end{center}

where $\lambda_0$, $\lambda_1$ and $\lambda_2$ are hyperparameters that control the relative importance of content loss and adversarial loss, respectively.

The variation in content loss ratios significantly influenced the model's performance. When the content loss ratio was reduced to zero, the model was able to focus more on low-frequency components, leading to an improvement in PSNR and SSIM values across the datasets. For instance, with an input resolution of 128$\times$128, setting the content loss ratio to zero resulted in a noticeable increase in performance, as the model could better optimize for low-frequency details. In contrast, when a content loss ratio was included, it forced the training slightly toward including more high-frequency components. However, the WaveMixSR-V2 architecture struggled to learn these components effectively, leading to sub-optimal performance compared to the scenario where content loss was not used.

\subsubsection{Gaussian Noise}
Regarding the addition of Gaussian noise, the results were mixed. An improvement was observed when using an input resolution of 64$\times$64, where the PSNR and SSIM values increased, suggesting that the noise channel helped the model capture finer details and potentially enhance the perceptual quality of the images. However, when we increased the input resolution to 128$\times$128, the inclusion of noise led to a slight decrease in performance. This indicates that the effectiveness of adding noise may depend on the input resolution and the model's ability to utilize this additional information effectively.

Interestingly, regular training of WaveMixSR-V2 using just PSNR as the pixel loss provided considerably better performance than GAN training. This could be due to the fundamental difference in how WaveMixSR-V2 and GANs handle frequency components in images. While GAN training typically forces the network to include more high-frequency details to enhance visual fidelity, WaveMixSR-V2 focuses more on low-frequency components. This divergence in focus likely led to a mismatch during training, making it challenging for the model to converge effectively under the GAN training. Consequently, the benefits of using GAN training with WaveMixSR-V2 were limited, as the architecture's emphasis on low-frequency components did not align well with the GAN's objectives.

\end{document}